# Value-at-Risk and backtesting with the APARCH model and the standardized Pearson type IV distribution

S. Stavroyiannis[(a,*)]

## Abstract

We examine the efficiency of the Asymmetric Power ARCH (APARCH) model in the case where the residuals follow the standardized Pearson type IV distribution. The model is tested with a variety of loss functions and the efficiency is examined via application of several statistical tests and risk measures. The results indicate that the APARCH model with the standardized Pearson type IV distribution is accurate, within the general financial risk modeling perspective, providing the financial analyst with an additional skewed distribution for incorporation in the risk management tools.

[a] Department of Accounting and Finance, School of Management and Economics, Technological Educational Institute of Peloponnese.
[*] e-mail: computmath@gmail.com
tel: (+30)2721045303
fax: (+30)2721045151

## 1.    Introduction

The recent global financial crisis, what came to be known as the subprime mortgage crisis, initiated an era of bank failures, credit squeezes, private defaults and massive layoffs. Under this perspective, financial time series modeling is of outmost importance due to the wide range of risk factors associated with the enormous growth of trading activity that has been taking place, often leading to an increase in financial uncertainty and volatility in the stock market returns. Financial time series data analysis differs from other time series analysis because the financial theory and its empirical time series contain an element of complex dynamic system with a high volatility and a great amount of noise, making the series to exhibit some statistical regularity, which are known as stylized facts. These include volatility clustering, the leptokurtosis effect, the leverage effect, skewness, long-range dependence in the data, and the long-run memory effect. Most of these stylized facts of financial time series, as well as Value-at-Risk (VaR) and Expected Shortfall (ES) measures, are consistently modeled within the Generalized Autoregressive Conditional Heteroskedasticity (GARCH) models framework of Engle and Bollerslev [1-2], and parsimonious families of GARCH models include the GJR-GARCH [3], the Exponential GARCH [4], and the APARCH [5] models. A typical approach is, after modeling the conditional mean and the conditional variance the parameters are estimated via maximization of the log-likelihood function (MLE), assuming some data generation process for the residuals. Although there is large variety of GARCH models, few continuous distributions have been used in financial time series modeling. These include the standard normal distribution applied by Engle [1], the Student-t distribution applied by Bollerslev [6], the Generalized Error Distribution (GED) introduced by Subbotin [7] and applied by Nelson [4], and the skewed t-Student distribution in the form of Fernandez and Steel [8] and applied by Lambert and Laurent [9].

The aim of this work is to elaborate on the properties of the standardized Pearson type IV (SPIV) distribution, and after introducing the case of the APARCH model [5] where the residuals follow the SPIV distribution, to examine the performance of the model via a variety of loss functions, VaR and ES tests, and backtesting measures. Section 2 reviews the literature of the Pearson type IV distribution, Section 3 describes the data and the model used, Section 4 provides an application, and Section 5 offers the concluding remarks.

## 2.    Review of the Pearson type IV distribution

In an attempt to construct a system of probability distributions for application to datasets in which the skewness and kurtosis deviated from the Gaussian distribution, Karl Pearson [10-13] classified seven types (I-VII) of distributions where the skewness and the kurtosis could also be fitted, while some extra classes (IX-XII) were also included [13]. The Pearson system of distributions is obtained by generalization of the differential equation leading to the Gaussian distribution, to the differential equation

$$\frac{1}{p(x)}\frac{dp(x)}{dx} = \frac{m-x}{a+bx+cx^2},$$  (1)

with the solution for the probability density function (PDF)

$$p(x) = (a+bx+cx^2)^{-\frac{1}{2c}} \exp\left(\frac{b+2cm}{c\sqrt{4ac-b^2}}\tan^{-1}\left(\frac{b+2cx}{4ac-b^2}\right)\right).$$  (2)

The Pearson system incorporates a variety of distributions; the normal distribution (type -0), the Beta (type I), the continuous uniform distribution (limit of the type I), the chi-squared, Gamma, and exponential distributions (type III), the Cauchy (or Lorentz, or Breit-Wigner) distribution (limit of the type IV), the inverse Gamma, and the inverse chi-squared distributions (type V), the F-distribution (type VI), the t-Student location scale distribution (type VII), and the monotonically decreasing power distribution (type VIII). Recent thorough reviews are provided by Magdalinos and Mitsopoulos [14], and Jondeau et al. [15]. If the discriminant $b^2-4ac$ is negative, after rearrangement of the terms in Eq. (2) we conclude on the Pearson type IV distribution in its recent form

$$p(x) = k\left[1+\left(\frac{x-\lambda}{a}\right)^2\right]^{-m}\exp\left[-\nu\tan^{-1}\left(\frac{x-\lambda}{a}\right)\right],$$  (3)

where, $k$ is the normalization constant

$$k = \frac{2^{2m-2}\left|\Gamma(m+i\nu/2)\right|^2}{\pi a \Gamma(2m-1)} = \frac{\Gamma(m)}{\sqrt{\pi}\,a\Gamma(m-1/2)}\left|\frac{\Gamma(m+i\nu/2)}{\Gamma(m)}\right|^2,$$  (4)

$\Gamma(\cdot)$, is the Gamma function, and $i$ is the imaginary unit. The parameters and their constraints are as follows; $\lambda$ is the location parameter, $a>0$ is the scale parameter, $\nu$ accounts for the skewness of the distribution, and $m>1/2$ (so that the normalization coefficient exists), accounts for the kurtosis of the distribution. Pearson [11] and Nagahara [16-18] use the following approach to work with the ratio of the complex Gamma functions

$$\left|\frac{\Gamma(x+iy)}{\Gamma(x)}\right|^2 = \prod_{n=0}^{\infty}\left[1+\left(\frac{y}{x+n}\right)^2\right]^{-1},\tag{5}$$

which in return was proven to be computationally intensive and not that accurate, even when only moderate precision is required. This led to fitting a dataset with the Pearson type IV distribution using the method of moments, where the normalization constant is not involved in the calculations [19-23]. Heinrich [24] noticed that

$$\left|\frac{\Gamma(x+iy)}{\Gamma(x)}\right|^2 = \frac{1}{{}_2F_1(-iy,iy;x;1)},\tag{6}$$

where ${}_2F_1(\cdot)$ is the Gauss hypergeometric function (GHF), and proposed a workable strategy to calculate the hypergeometric series; reconstruct Eq. (6) using the relation $\Gamma(z+1) = z\Gamma(z)$, and start from a large $n$ working down to $n = 0$

$$\left|\frac{\Gamma(x+iy)}{\Gamma(x)}\right|^2 = \left[1+\left(\frac{y}{x}\right)^2\right]^{-1}\left|\frac{\Gamma(x+1+iy)}{\Gamma(x+1)}\right|^2 = \cdots.\tag{7}$$

In the same work [24] the cumulative distribution function (CDF), which is needed for the calculation of the constants at the confidence intervals is also computed

$$P(x) = p(x)\frac{2}{2m-1}\left(i-\frac{x-\lambda}{a}\right){}_2F_1\left(1, m+i\nu/2; 2m; \frac{2}{1-i(x-\lambda)/a}\right),\tag{8}$$

where, $p(x)$ is the PDF given in Eq. (3). Although the formula appears to be complex, in the end the result is real after cancelation of the imaginary terms in the series summation.

One drawback of the parameterization of the Pearson type IV distribution in Eq. (3) is that $\lambda$ and $a$ are the conditional mode and some measure of conditional dispersion, and not the conditional mean and the conditional variance. In order to keep the usual econometric tradition it is important to express the probability density in terms of the mean and the variance therefore, a re-parameterization is needed, aiming to get a Pearson type IV distribution with zero mean and unit variance, to preserve the martingale hypothesis used in financial time series [9]. Such attempts, implementing constant and dynamic conditional skewness and kurtosis have been performed in [25-31]; however, in most of the cases the coefficients $\lambda$ and $a$ are included in the PDF and the maximum likelihood estimation (MLE). An econometric model with re-parametrization of Eq. (3) resulting in the standardized form was proposed in [32-33], including conventional in-sample and out-of-sample VaR tests.

## 3. Econometric methodology

### 3.1. The data

We consider the daily returns of Cushing Oklahoma Crude oil, West Texas Intermediate (WTI) spot price, Free on Board (FOB), in Dollars per Barrel, from Apr-2-1990 to Sep-28-2015, and the return is realized via the successive differences of the natural logarithm of the close prices, $r_t = \ln(p_t / p_{t-1}) \times 100$. Only weekdays are used and in case of unavailable data because of national holidays, bank holidays, or any other reasons, the previous close value is considered. The series is publically available from the Economic Research Division of the Federal Reserve Bank of St. Louis, for future models and results comparison.

## 2. The model

The dynamics of the APARCH model is expressed as follows:

$$r_t = \mu + \varepsilon_t = \mu + \sigma_t z_t, \tag{9}$$

$$\sigma_t^\delta = \omega + ak(\varepsilon_{t-1})^\delta + \beta \sigma_{t-1}^\delta, \tag{10}$$

$$k(\varepsilon_{t-1}) = |\varepsilon_{t-1}| - \gamma \varepsilon_{t-1}, \tag{11}$$

where $a$ and $\beta$ are the ARCH and GARCH coefficients, $\gamma$ is the leverage effect capturing the leverage or asymmetry effect in return volatility [34], and $\delta$ is the Taylor effect [35] (Taylor, 1986) regarding the difference in the sample autocorrelations of absolute and squared returns. The APARCH model is a nested model including as special cases the ARCH ($\delta = 2, \beta = 0, \gamma = 0$), the GARCH ($\delta = 2, \gamma = 0$), the TS-GARCH ($\delta = 1, \gamma = 0$), the GJR ($\delta = 2$), the TARCH ($\delta = 1$), the NARCH ($\gamma = 1, \beta = 0$) and the Log-ARCH ($\delta \to 0$) [36]. The residuals follow the SPIV distribution as defined in [32-33], where PDF is

$$p(z) = \frac{\tilde{\sigma}\Gamma\left(\frac{m+1}{2}\right)}{\sqrt{\pi}\Gamma\left(\frac{m}{2}\right)} \left| \frac{\Gamma\left(\frac{m+1}{2} + i\frac{\nu}{2}\right)}{\Gamma\left(\frac{m+1}{2}\right)} \right|^2 \frac{\exp(-\nu \tan^{-1}(\hat{\sigma}z + \hat{\mu}))}{\left(1 + (\hat{\sigma}z + \hat{\mu})^2\right)^{\frac{m+1}{2}}}, \tag{12}$$

$$\hat{\mu} = -v\,/(m-1)\,, \tag{13}$$

$$\hat{\sigma} = \sqrt{1/(m-2)\left(1 + v^2\,/(m-1)^2\right)}\,, \tag{14}$$

and the log-likelihood is,

$$L_{PIV} = N \ln C - \sum_{t=1}^{N}\left[1/2\ln h_t + \frac{m+1}{2}\ln(1 + (\hat{\sigma}z_t + \hat{\mu})^2) + v\tan^{-1}(\hat{\sigma}z_t + \hat{\mu})\right], \tag{15}$$

$$C = \Gamma\!\left(\frac{m+1}{2}\right)\!/\Gamma\!\left(\frac{m}{2}\right) + 1/2\ln(\hat{\sigma}) - 1/2\ln(\pi) + \frac{1}{2}\ln\left|\Gamma\!\left(\frac{m+1}{2} + i\frac{v}{2}\right)\!/\Gamma\!\left(\frac{m+1}{2}\right)\right|. \tag{16}$$

The constraints used for the optimization are $a > 0$, $\beta \geq 0$, $\delta > 0$, and $-1 < \gamma < 1$, while for the existence of a stationary solution the constraint $\alpha \mathrm{E}\left(\left|z_t\right| - \gamma z_t\right)^\delta + \beta < 1$ has to be included. Following Laurent [37] it is convenient to start the recursion of Eq. (10) by setting the unobserved components to their sample average; $k(\varepsilon_{t-1})^\delta = (1/T)\sum_{s=1}^{T}\left(\left|\varepsilon_{t-1}\right| - \gamma\varepsilon_{t-1}\right)^\delta$, for $t \leq 1$, and $\sigma_t^\delta = \left((1/T)\sum_{s=1}^{T}\varepsilon_s^2\right)^{\delta/2}$, for $t \leq 0$. All calculations thereinafter have been performed with native code with the Matlab® computing language.

## 3.    Results and discussion

### 3.1.    Stylized facts

Fig. 1 shows the time series under consideration (top left), the returns (top right), the autocorrelation function (ACF) (bottom left) and the partial autocorrelation function (PACF) (bottom right) of the returns for 12 lags.

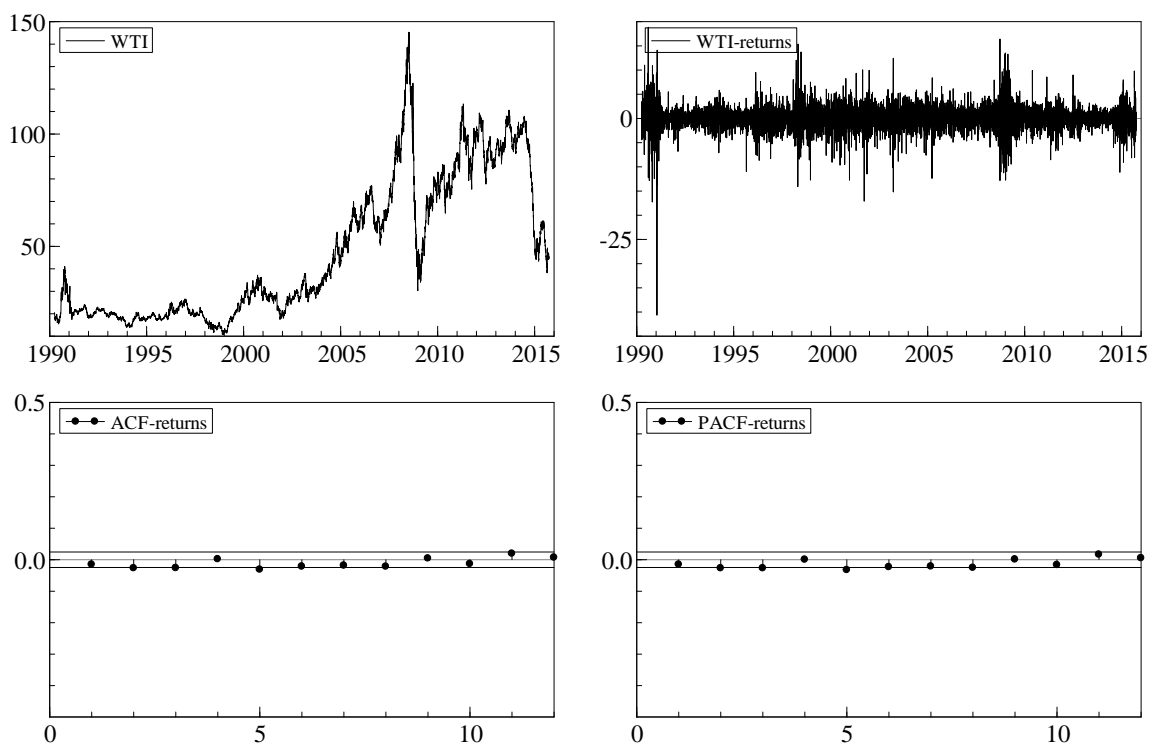

**Fig. 1** time series under consideration (top left), the returns (top right), the ACF (bottom left) and the PACF (bottom right) of the returns for 12 lags.

The WTI price and returns show the leverage or asymmetry effect, that stock market volatility increases with bad news and decreases with good news, or that stock market volatility tends to rise following negative returns and fall following positive returns. The summary statistics of the WTI returns series are presented in Table 1. Specifically, we report information on the minimum and maximum values, the range, and the four moments of the data. The returns are negatively skewed and leptokurtic as revealed by the value of the Jarque–Bera (J.B.) normality test. The Ljung–Box (LB) test for 12 lags indicates serial correlation in both the returns and the squared returns (LB-sq.), and the ARCH test reveals heteroskedasticity. The results of the two tests, the augmented Dickey-Fuller (ADF) and the Kwiatkowski, Phillips, Schmidt, and Shin (KPSS), indicate the absence of unit root in the returns series. The star (*) notation thereinafter indicates statistical significance at the 5% and 1% critical levels.

**Table 1** Stylized facts of the WTI returns

| Commodity | WTI |
|-----------|---------|
| min | -40.64 |
| max | 18.868 |
| range | 59.508 |
| mean | 0.0116 |
| std. dev. | 2.4467 |
| skewness | -0.8208 |
| kurtosis | 19.7660 |
| J.B. | 78630.0 |
| ARCH(12) | 29.418* |
| LB(12) | 30.0661* |
| LB(12)-sq | 545.67* |
| ADF | -49.472* |
| KPSS | 0.0749 |

In the empirical literature, asset returns are commonly found to be approximately uncorrelated over time, while non-linear transformations such as powers of absolute returns $\left| r_t \right|^p$ and their logarithms show significant autocorrelation for many lags. This was first noted by Taylor [35] (1986), who found that for various financial series the autocorrelations are higher for the absolute returns than for the squared ones. Further studies on stock indices and exchange rates by [38-39] Ding and Granger (1996) and Granger, Ding, and Spear (1997) found the autocorrelations of $\left| r_t \right|^p$ to be highest for value $p = 1$ but also for smaller values $p < 1$, while the autocorrelations of $\left| r_t \right|$ were always found to be bigger than those of $\left| r_t \right|^2$ in the aforementioned studies, leading Malmsten and Teräsvirta [40] to term this observation as the Taylor effect. The ACF and the PACF of the absolute and squared returns are shown in Fig. 2 indicating the presence of the Taylor effect.

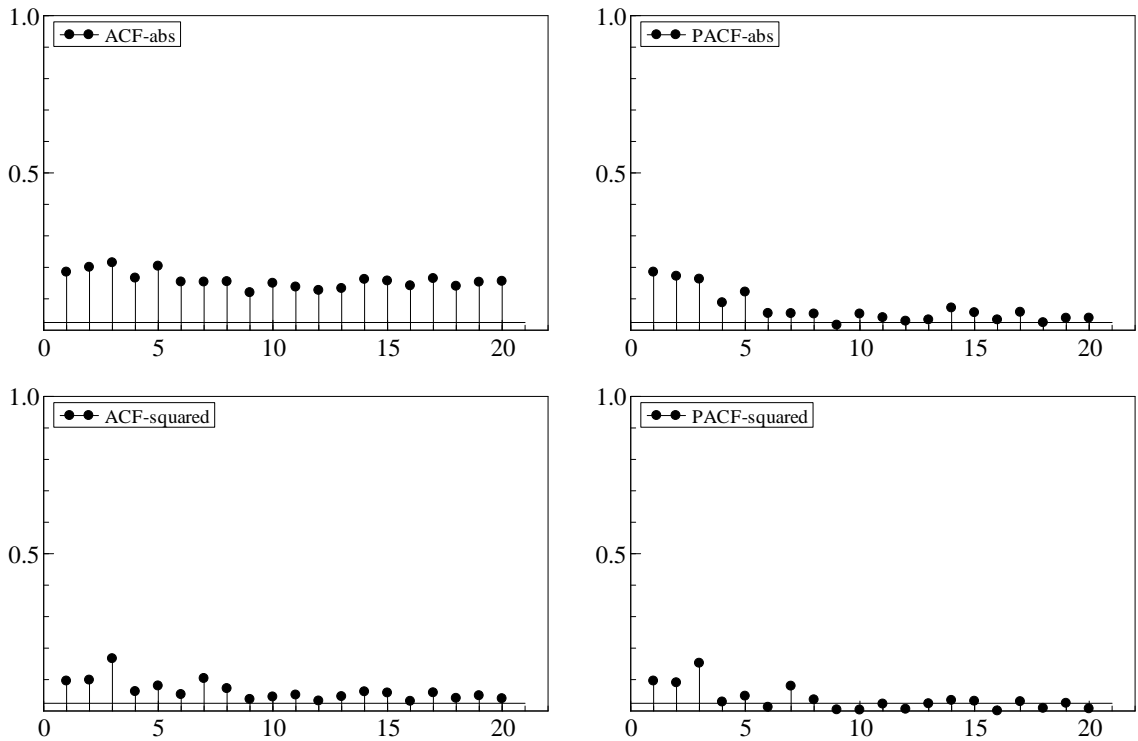

**Fig.2** The ACF (left) and the PACF (right) for the absolute returns (top) and squared returns (bottom).

### 3.2. Estimation results and loss functions

The results of the optimization are shown in Table 2 where except the constant in mean all other coefficients are statistically significant. The GARCH parameters that is the error coefficient $a$ which is associated with the reaction of the volatility to market movements, and $\beta$ which is associated with the time a market shock takes to decrease, are both statistical significant. The leverage parameter $\gamma$, and the power of the conditional heteroskedasticity $\delta$ are statistically significant in favor of the leverage and Taylor effects. The statistically significant deviation of the value of the fitted parameter $\delta$ from the value 2 shows the strong and persistent autocorrelation of the $\left|r_t\right|^p$ function, which is an indication for the long memory property of the returns. The value of the Log-Likelihood estimator using the SPIV distribution (-14132.79) is an improvement over most distributions used in econometric software that is, the Normal (-

14409.48), the GED (-14156.48), the Student location-scale (-14141.32), and the Skewed t-Student (-14135.90), using the same APARCH model.

**Table 2** Results of the APARCH(1,1)-Pearson IV model.

| Coefficient | Value | Robust t-stat | p-value |
|---|---|---|---|
| $\mu$ | 0.0058 | 0.0086 | 0.6745 |
| $\omega$ | 0.0166* | 0.0041 | 3.9988 |
| $\alpha$ | 0.0586* | 0.0057 | 10.2264 |
| $\beta$ | 0.9493* | 0.0050 | 187.9879 |
| $\gamma$ | 0.2043* | 0.0640 | 3.1901 |
| $\delta$ | 1.1946* | 0.1168 | 10.2288 |
| $v$ | 0.4748* | 0.1243 | 3.8212 |
| $m$ | 5.6275* | 0.4082 | 13.7853 |
| persistence | 0.9940 | | |
| Log-Lik. | -14132.79 | | |

The evaluation of volatility forecasts raises the problem that the variable of interest is latent, and this can be solved by replacing the latent conditional variance by a proxy. In a general framework, Hansen and Lunde [41] show that when the evaluation is based on a target observed with error, the choice of the loss function becomes critical in order to avoid a distorted outcome and provide conditions on the functional form of the loss function which ensure consistency of the proxy based ordering. However, such specific economic loss functions are rarely available, and purely statistical loss functions are more commonly used to evaluate both in-sample and out-of-sample volatility forecasts.

Letting the conditional variance to be $h_t = \sigma_t^2$ and the actual variance to be $\varepsilon_t^2$, since the squared innovations are used as a proxy for the conditional variance, the following loss functions have been considered in Table 3; the mean squared absolute error (MAE), the mean absolute deviation (MAD), the median absolute error (MedAE) and the median absolute percentage error (MedAPE), the heteroscedasticity adjusted MSE (HMSE) of Bollerslev and Ghysels [42] (1996), the heteroscedasticity adjusted MAE (HMAE) of Andersen, Bollerslev and Lange [43] (1999), the logarithmic loss (LL) function of Pagan and Schwert [44] (1990), and the loss function implicit in the Gaussian quasi-maximum likelihood function (GMLE) of Bollerslev, Engle and Nelson [45].

**Table 3** Results of the Loss functions

| Loss function | Formula | Value |
|---|---|---|
| mean squared absolute error (MSE) | $MSE = 1/N \sum_{t=1}^{N} (\varepsilon_t^2 - h_t)^2$ | 73.1299 |
| mean absolute deviation (MAD) | $MAD = 1/N \sum_{t=1}^{N} \left| \varepsilon_t - \sqrt{h_t} \right|$ | 2.5776 |
| median absolute error (MedAE) | $MedAE = median \left| h_t - \varepsilon_t^2 \right|$ | 6.5338 |
| median absolute percentage error (MedAPE) | $MedAPE = median \left( \left| h_t - \varepsilon_t^2 \right| / \varepsilon_t^2 \right)$ | 3.2798 |
| heteroscedasticity adjusted MSE (HMSE) | $HMSE = 1/N \sum_{t=1}^{N} \left( \varepsilon_t^2 / h_t - 1 \right)^2$ | 1.9902 |
| heteroscedasticity adjusted MAE (HMAE) | $HMAE = 1/N \sum_{t=1}^{N} \left| \varepsilon_t^2 / h_t - 1 \right|$ | 6.8624 |
| logarithmic loss (LL) | $LL = 1/N \sum_{t=1}^{N} \ln \left( \varepsilon_t^2 / h_t \right)$ | 1.0949 |
| Gaussian maximum likelihood (GMLE) | $GMLE = 1/N \sum_{t=1}^{N} \left( \ln(h_t) + \varepsilon_t^2 / h_t \right)$ | 12.6031 |

### 3.3. The cumulative distribution function

In order to perform VaR tests and backtesting the constants at the confidence levels are required, and a fast approach is the numerical integration of the PDF [46], or the employment in the calculations for the ghf of the Euler integral representation,

$$_2F_1(a,b;c;w) = \frac{\Gamma(c)}{\Gamma(b)\Gamma(c-b)} \int_0^1 t^{b-1}(1-t)^{c-b-1}(1-wt)^{-a} dt , \tag{17}$$

which for $\mathrm{Re}(c) > \mathrm{Re}(b) > 0$ holds for all $w$ in the complex plane cut along the real axis from 1 to $\infty$. From the analytical point of view the cdf of the SPIV distribution of Eq. (12) is given by,

$$P(z) = p(z) \frac{i - (\hat{\sigma}z + \hat{\mu})}{\hat{\sigma}m} \, _2F_1 \left( 1, \frac{m+1}{2} + i\frac{\nu}{2}; m+1; \frac{2}{1 - i(\hat{\sigma}z + \hat{\mu})} \right). \tag{18}$$

The GHF converges absolutely when the argument $|w| < 1$, and has a singularity at $w = 1$. Since form Eq. (13) $\mathrm{Re}(m+1-1-(m+1)/2) = (m+1)/2 > 1$, the GHF is well defined and converges absolutely on the unit circle. There is a branch cut associated with the singularity and by convention is chosen to lie on the real axis along with $\mathrm{Re}(w) > 1$. Therefore, for $z < -\hat{\mu}/\hat{\sigma} - \sqrt{3}/\hat{\sigma}$ the GHF in Eq. (18) is absolutely convergent. There is an interference with the branch cut if $z > -\hat{\mu}/\hat{\sigma} + \sqrt{3}/\hat{\sigma}$ and the simplest way is to apply the transformation,

$$P(z \mid m, \nu, \hat{\sigma}, \hat{\mu}) \equiv 1 - P(-z \mid m, -\nu, \hat{\sigma}, -\mu) , \tag{19}$$

and for the other two regions where $\left| z + \hat{\mu}/\hat{\sigma} \right| < \sqrt{3}/\hat{\sigma}$, one of the existing transformations [47-48] i.e. $z \to 1/z$ can be used,

$$P(z) = p(z)\frac{i}{\hat{\sigma}(m - i\nu - 1)}\left[1 + (\hat{\sigma}z + \hat{\mu})^2\right]_2 F_1\left(1, 1 - m; (3 - m)/2 + i\nu/2; \left[1 + i(\hat{\sigma}z + \hat{\mu})\right]/2\right)$$
$$+ \frac{1}{1 - \exp\left[-\pi(\nu + i(m + 1))\right]}. \qquad (20)$$

It should be noted that analytical calculations via Eqs. (18-20) are time consuming, and the actual numerical error compared to numerical integration or using the Euler representation of Eq. (17) is $O(10^{-10})$.

### 3.4.   VaR and backtesting

After the estimation of the parameters of the model the VaR for the short and long position is calculated via $VaR(a) = \mu + P^{-1}(a)\sigma$, where $a = 1 - c$ is the VaR level, $c$ is the confidence level, and $P^{-1}(\cdot)$ is the inverse of the CDF function at the specific VaR level. A typical VaR for the short and long position is shown in Fig. 3 for $a = 0.05$ and when an observation exceeds the VaR border it is called a VaR violation, or VaR break.

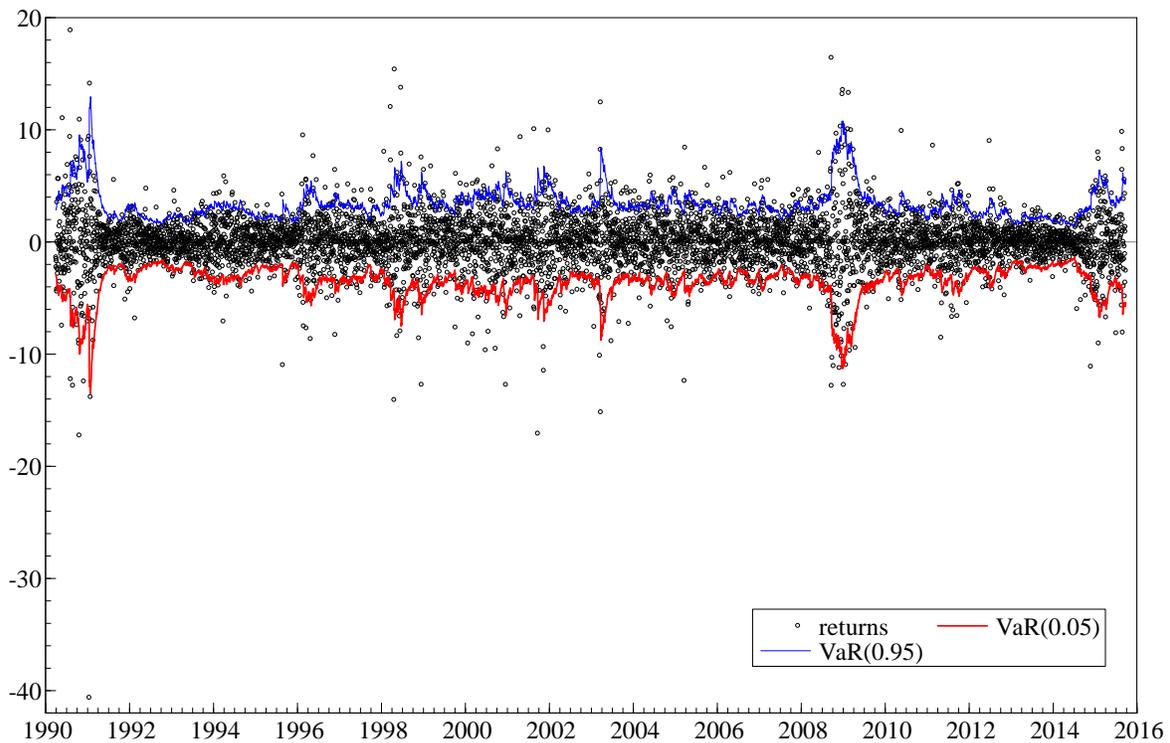

**Fig. 3** WTI returns (open circles), and Value-at-Risk positioning for the 0.05 (red line, bottom) and 0.95 (blue line, top) levels.

One way to examine the accuracy of a model is to count the number of VaR violations the model produces which has to be as close as possible to the number of VaR violations specified by the confidence level (Table 4, Success/Failure ratio). Since rarely the exact amount suggested by the confidence level is observed it comes down to whether the number of violations is reasonable or not, before a model is accepted or rejected. The most widely known test is the Kupiec test [49] based on the proportion of failure (POF), which measures whether the number of violations is consistent with the confidence level, under the null hypothesis that the model is correct (Table 4, POF p-value). In order to check whether the violations are spread evenly over time or they form clustering, the Christoffersen interval forecast test is used [50] conducted as a likelihood ratio test proportional to a $\chi^2(1)$ distribution (Table 4, Independence test p-value). Joining the two criteria, the Kupiec test and the Christoffersen independence test, results in the Christoffersen conditional coverage test [51] which provides a means to check in which regard the violation series fails the correct conditional coverage property, conducted as a likelihood ratio test proportional to a $\chi^2(2)$ distribution (Table 4, Conditional test p-value).

The conditional coverage test checks for clustering but it only uses consecutive data points; therefore, it only tests the clustering of one lag. Engle and Manganelli [52] suggested the dynamic quantile test (DQ-test), using a linear regression model linking current violations to past violations, so as to test the conditional efficiency hypothesis. Whatever the chosen specification, the null hypothesis test of conditional efficiency corresponds to testing the joint nullity of the test's coefficients. Therefore, if the parameters of the test are zero, the current VaR violations are uncorrelated to past violations, as a consequence of the independence hypothesis, whereas the unconditional coverage hypothesis is verified if the constant term is also zero. The test is conducted via Wald statistics which is proportional to a $\chi^2(2K+1)$ distribution, where $K$ is the number of lags included in the regression, and in this work $K = 5$ (Table 4, DQ test p-value).

Two more tests not incorporated in software and rarely seen in the literature are the Lopez and Sarma approaches. Lopez (1999) [53] suggested the development of a loss function for back-testing different models and proposed to measure the accuracy of the VaR forecasts on the basis of the distance between the observed returns $r_t$, and the forecasted VaR values if a violation occurs

$$L_t = \begin{cases} 1 + (r_t - VaR_t)^2, \text{ if } r_t < VaR_t \\ 0, \qquad\qquad\quad \text{ if } r_t \geq VaR_t \end{cases}. \tag{21}$$

A VaR model is penalized when an exception takes place. Hence, the model is preferred to other candidate models if it yields a lower total loss value. This is defined as the sum of these penalty scores. This function incorporates both the cumulative number of exceptions and their magnitude. However, a model that does not generate any violation is deemed the most adequate since the sum is zero. Thus, the risk models must be first filtered by using the aforementioned back-testing measures too. The results of the Lopez measure are shown in Table 4 (Lopez).

Sarma et al. (2003), [54] combining the advantages of a loss function with those of back-testing measures, suggested a two-stage back-testing procedure. When multiple risk models meet the back-testing statistical criteria of VaR evaluation, a loss function is brought into play to judge statistically the differences among VaR forecasts. In the first stage, the statistical accuracy of the models is tested by examining whether the mean number of violations is not statistically significantly different from that expected and whether these violations are independently distributed. In the second stage, they propose use of what they term the firm's loss function, i.e., penalizing failures but also imposing a penalty reflecting the cost of capital suffered on other days, the regulatory loss function

$$L_t = \begin{cases} (r_t - VaR_t)^2, \text{if } r_t < VaR_t \\ 0, \qquad\quad \text{if } r_t \geq VaR_t \end{cases}, \tag{22}$$

and firm's loss function where a represents the cost of opportunity capital

$$L_t = \begin{cases} (r_t - VaR_t)^2, \text{if } r_t < VaR_t \\ -aVaR_t, \qquad \text{if } r_t \geq VaR_t \end{cases}. \tag{23}$$

The results for the regulatory loss function are shown in Table 4 (Sarma). The Sarma approach addresses the different conceptual approach between the regulators and the risk managers regarding the aiming of the market risk management tool. Regulators are interested in the number of VaR breaks and the size of the non-covered losses, while risk managers disagree on safety and profit maximization, since an excessively high VaR forces them to hold too much capital, imposing the opportunity cost of capital upon firms.

**Table 4** Results of the VaR and expected shortfall tests

| Quantile | Success ratio | Kupiec POF p-value | Independence test p-value | Conditional test p-value | DQ-test p-value | Lopez | Sarma |
|---|---|---|---|---|---|---|---|
| 0.9500 | 0.950376 | 0.888000 | 0.921350 | 0.985320 | 0.87675 | 1899.053 | 1569.053 |
| 0.9750 | 0.976241 | 0.513550 | 0.123730 | 0.247050 | 0.16032 | 1099.768 | 941.7679 |
| 0.9900 | 0.991128 | 0.345980 | 0.554570 | 0.538670 | 0.080552 | 573.4669 | 514.4669 |
| 0.9950 | 0.995489 | 0.565570 | 1.000000 | 0.847830 | 0.36923 | 351.6168 | 321.6168 |
| 0.9975 | 0.997143 | 0.568590 | 1.000000 | 0.850000 | 0.008431 | 209.3916 | 190.3916 |
| 0.9990 | 0.998647 | 0.387180 | 1.000000 | 0.688060 | 0.995 | 93.592 | 84.592 |
| Quantile | Failure ratio | Kupiec POF p-value | Independence test p-value | Conditional test p-value | DQ-test p-value | Lopez | Sarma |
| 0.0500 | 0.049323 | 0.799700 | 0.091550 | 0.233260 | 0.037659 | 3373.896 | 3045.896 |
| 0.0250 | 0.023459 | 0.416010 | 0.495950 | 0.569740 | 0.68114 | 2274.476 | 2118.476 |
| 0.0100 | 0.009474 | 0.663420 | 0.630770 | 0.810420 | 0.86339 | 1458.575 | 1395.575 |
| 0.0050 | 0.005113 | 0.896640 | 1.000000 | 0.991600 | 0.98866 | 1079.976 | 1045.976 |
| 0.0025 | 0.003308 | 0.208720 | 1.000000 | 0.453780 | 0.94595 | 825.706 | 803.706 |
| 0.0010 | 0.001203 | 0.611860 | 1.000000 | 0.879210 | 0.9996 | 566.3737 | 558.3737 |

## 3.5.  Expected Shortfall and Tail Conditional Expectation

ES emerged as a natural alternative to Value-at-Risk fulfilling all four axioms of a coherent risk measure [55-56], and belongs to the category of spectral risk measures which are not elicitable unless they reduce to minus the expected value of the losses conditional on the loss being larger than the VaR [57-58], which is known as the Tail Conditional Expectation (TCE), or Conditional VaR (CVaR). Most software report on TCE (Table 5, TCE1), and another measure as indicated by Hendricks [59] is the TCE1 divided by the associated VaR values (Table 5, TCE2). The TCE2 measure reports on the degree to which events in the tail of the distribution typically exceed the VaR measure, by calculating the average multiple of these outcomes to their corresponding VaR measures. The TCE is a coherent risk measure only when restricted to continuous elliptical distribution functions, while it may violate subadditivity on general distributions [57-58], or when used in historical data calculations. In such a case a correction has to be made and a compact expression useful in numerical computations, connecting the ES, TCE, and VaR, is given by Rockafellar and Uryasev [60]

$$ES^{(a)}(X) = TCE^{(a)} + (\lambda - 1)(TCE^{(a)} - VaR^{(a)}), \tag{24}$$

where, $\lambda$ is the probability on the loss being larger than the VaR, divided by the VaR level. The results on the VaR, and ES are shown in Table 5 (VaR, ES).

**Table 5** Results of the VaR and expected shortfall tests

| Quantile | VaR | TCE1 | TCE2 | ES |
|---|---|---|---|---|
| 0.9500 | 1.5356 | 4.8593 | 1.3724 | 4.8593 |
| 0.9750 | 1.9186 | 5.7987 | 1.3253 | 5.7987 |
| 0.9900 | 2.4472 | 7.5678 | 1.3267 | 7.5678 |
| 0.9950 | 2.8771 | 9.0827 | 1.3408 | 9.0827 |
| 0.9975 | 3.3424 | 10.406 | 1.2886 | 11.4148 |
| 0.9990 | 4.0262 | 11.483 | 1.2508 | 14.1187 |
| Quantile | VaR | TCE1 | TCE2 | ES |
| 0.0500 | -1.6187 | -5.1431 | 1.4353 | -5.1431 |
| 0.0250 | -2.0707 | -6.4100 | 1.3927 | -6.4100 |
| 0.0100 | -2.7090 | -8.5625 | 1.3716 | -8.5625 |
| 0.0050 | -3.2373 | -10.755 | 1.3584 | -10.9245 |
| 0.0025 | -3.8161 | -11.096 | 1.2906 | -13.4496 |
| 0.0010 | -4.6757 | -15.104 | 1.4095 | -17.2211 |

## 4.    Conclusion

In this work we have examined the performance of the APARCH model in the case where the data generation process follows the SPIV distribution. In particular we are interesting in combining a parsimonious and flexible GARCH model that takes into account most of the stylized facts of financial time series, with the SPIV distribution. The model's log-likelihood function value is improved compared to most distributional schemes available in software, denoting that it captures the data generation process more reliably. We apply a variety of loss functions and VaR tests that can be used for future comparison with other distributions, and the results indicate that the model is accurate within the general financial risk modeling perspective. The SPIV distribution has not attracted much interest in the literature due to the mathematical and computational difficulty. Despite the fact that due to the mathematical and computational difficulty the SPIV distribution has not attracted much interest in the literature, it might provide the financial analyst with an additional distributional scheme to be used in econometric modeling and financial risk management.